\documentclass[prd,aps,twocolumn,showpacs,preprintnumbers,amsmath,nofootinbib,amssymb]{revtex4}
\usepackage[dvips]{graphicx} 
\usepackage{amsmath} 
\usepackage{amssymb} 
\usepackage{verbatim} 
\usepackage{graphicx}
\usepackage{dcolumn}
\usepackage{bm}
\def\be{\begin{equation}} 
\def\ee{\end{equation}} 
\def\bea{\begin{eqnarray}} 
\def\eea{\end{eqnarray}} 
 
\begin{document} 
 
 
\date{\today} 
 
\title{The 21~cm Signature of Cosmic String Wakes} 
 
\author{Robert H. Brandenberger$^{1}$, Rebecca J. Danos$^{1}$, Oscar F. Hern\'andez$^{2,1}$ 
and Gilbert P. Holder$^{1}$
\email[email: ]{rhb,rjdanos,oscarh,holder@physics.mcgill.ca}}
 
\affiliation{1) Department of Physics, McGill University, 
Montr\'eal, QC, H3A 2T8, Canada \\
2) Marianopolis College, 4873 Westmount Ave., Westmount, QC H3Y 1X9, Canada}

 
\begin{abstract} 

We discuss the signature of a cosmic string wake in 21cm redshift surveys.
Since 21cm surveys probe higher redshifts than optical large-scale structure
surveys, the signatures of cosmic strings are more manifest in 21cm maps
than they are in optical galaxy surveys. We find that, provided the
tension of the cosmic string exceeds a critical value (which depends on
both the redshift when the string wake is created and the redshift of
observation), a cosmic string wake will generate an emission signal with
a brightness temperature which approaches a limiting value which at
a redshift of $z + 1 = 30$ is close to
400 mK in the limit of large string tension. The signal will have a
specific signature in position space: the excess 21cm radiation will be
confined to a wedge-shaped region whose tip corresponds to the
position of the string, whose planar dimensions are set by the
planar dimensions of the string wake, and whose thickness 
(in redshift direction) depends on the string tension. For wakes
created at $z_i  + 1 = 10^3$, then at a redshift
of $z + 1 = 30$ the critical value of the string tension $\mu$ is
$G \mu = 6 \times 10^{-7}$, and it decreases linearly with redshift
(for wakes created at the time of equal matter and radiation, 
the critical value is a factor of two lower at the same redshift).
For smaller tensions, cosmic strings lead to an observable
absorption signal with the same wedge geometry.   

\end{abstract} 
 
\maketitle

\newcommand{\eq}[2]{\begin{equation}\label{#1}{#2}\end{equation}} 
 
\section{Introduction} 

Cosmic strings (see \cite{Zel,Vil,Kibble,TB,Sato,Stebbins} 
for initial work on cosmic strings and structure formation)
cannot be \cite{Albrecht,Turok}
the dominant source of the primordial fluctuations, however
they can still provide a secondary source of fluctuations.
Over the past decade, the realization has grown  that many inflationary 
scenarios constructed in the context of supergravity models lead to the 
formation of gauge theory cosmic strings at the end of 
the inflationary phase \cite{Rachel}. Also, in a large class of brane
inflation models the formation of cosmic superstrings \cite{Witten}
at the end of inflation is
generic \cite{CS-BI}, and in some cases  (see \cite{Pol1}) these
strings are stable (see also \cite{recentCS} for reviews on fundamental
cosmic strings). Cosmic superstrings are also a possible remnant
of an early Hagedorn phase of string gas cosmology \cite{SGrev}.
In all of these contexts, both a scale-invariant spectrum of
adiabatic coherent perturbations and a sub-dominant contribution
of cosmic strings is predicted. Hence, it is important to search
for the existence of cosmic strings. 

In this paper we study the signature of cosmic strings in 21cm 
radiation maps (see e.g. \cite{Furlanetto} for an in-depth review of
21cm cosmology).
Observing the intensity of the cosmological background radiation at
wavelengths corresponding to the red-shifted 21cm transition line of
neutral hydrogen has
several potential advantages compared to the currently explored
windows . First of all, it probes the distribution of the dominant
form of baryonic matter and is thus not sensitive to our incomplete
understanding of star formation and non-linear evolution, which
is a problem when interpreting the results of optical redshift
surveys. Secondly, it probes the universe at higher redshifts
and allows us to explore the ``dark ages" (the epoch before star
formation and non-linear clustering set in). Related to this, it
explores the distribution of matter in a regime
when the amplitude of the fluctuations is smaller and linear theory
is a better approximation. Finally, 21cm surveys provide
three-dimensional maps, a significant potential advantage
over CMB anisotropy maps.

Cosmic strings are known to give rise to distinctive signatures
in CMB temperature anisotropy maps \cite{KS}, CMB
polarization maps \cite{Danos3} and large-scale structure (LSS)
maps \cite{Silk,Rees,Vach,SVBST,Charlton,Hara}. These distinctive
signatures come from moving long (compared to the Hubble
radius) strings (see Section 2). In the CMB temperature maps,
these strings lead to line discontinuities, in the polarization
maps to (roughly) rectangular regions with extra polarization,
and in LSS maps to thin planar regions of enhanced density.
These signals are manifest in position space maps, but
they become obscured when calculating power spectra.
Hence, the lesson is to study the maps in position versus momentum space.

Here, we compute the signature of a single cosmic string
wake in 21cm emission. We find that strings with tensions $\mu$
somewhat below the current  limits of $G \mu = 3 \times 10^{-7}$
could be detected in 21cm maps where they would appear as
wedges in 21cm maps with either extra emission or extra
absorption, depending on the tension and on the 
specific redshift.
The planar dimensions are set by the direction of the string and
its velocity vector, and the width of the wedges is proportional
to the string tension. This signal must be searched for in
position space maps. In Fourier space the distinctive phase
information would be washed out.

The outline of this paper is as follows:  In Section 2 we discuss how
string wakes are generated and how these lead to
distinctive signals for observations. We review
gravitational accretion of matter onto cosmic string
wakes and compute the temperature of the HI gas
inside the wake. In Section 3 we then study the
21cm emission signal from a single cosmic string
wake. The generalization to the case of a network 
of string wakes will be addressed
in a future publication. We
compute the brightness temperature and describe
the geometrical structure of the signal, a structure
very characteristic for cosmic strings. In the final
section we summarize our results and put them in
the context of other work on the possible detection of
cosmic strings. When computing the magnitude of the
21cm signal, we use the same WMAP concordance values
\cite{WMAP} for the cosmological parameters as used in
the review \cite{Furlanetto}.

\section{Cosmic Strings and Large-Scale Structure}

In many field theory models, the formation of a network of
cosmic strings is an inevitable consequence of a phase
transition in the early universe (for reviews on cosmic strings 
see e.g. \cite{Kibble2,VilShell,HK,RHBrev4}).
This network of cosmic strings approaches
a ``scaling solution" which means that the statistical
properties of the string network are the same at all
times if distances are scaled to the Hubble radius \cite{RHBrev4} .
The detailed form of the scaling solution must be
obtained from numerical simulations  \cite{CSsimuls}
of the evolution of a network of cosmic strings (see also
\cite{Pol2} for some recent analytical work).
There are two components of the string network -
firstly a network of ``long"  strings with a mean curvature
radius $\zeta = c_1 t$, where $c_1$ is a constant
of order unity, and secondly a distribution of
string loops with radii smaller than the horizon
which results from the ``cutting up" of the long
string network as a consequence of string
intersections. According to more recent cosmic
string evolution simulations, the long string
component is more important for cosmological
structure formation. 

Numerical simulations
by different groups have clearly verified the
scaling solution for the long string network.
There is, however, still a large uncertainty
concerning the distribution of string loops (the
only agreement seems to be that the loops
are less important than the long strings for
cosmological structure formation). Hence, in
order to obtain constraints in models with
cosmic strings which are robust against
the uncertainties in the numerical simulations, it
is important to focus on signatures of long strings
as opposed to signatures of string loops.
The tightest current constraints on cosmic
strings come from the shape of the angular
power spectrum of CMB anisotropies,
yielding a constraint of \cite{Wyman} (see
also \cite{others})
\be \label{limit}
G \mu \, < \, 3 \times 10^{-7} \, ,
\ee
where $\mu$ is the string tension and $G$ is Newton's constant. 
However, it is important to keep in mind
that this limit is based on an analytical
description of the scaling solution which
contains a number of parameters which 
can only be determined from comparisons
with numerical simulations and whose values
thus have a substantial uncertainty. Work
based on a field theory simulation of strings
\cite{Hind} gives a limit of twice the above
value (see also \cite{recent} for a very recent analysis
of the different bounds). Note that limits on the cosmic string
tension which come from gravitational
radiation from string loops \cite{CSgrav} are
sensitive both to the large uncertainties in
the distribution of string loops, and also
to back-reaction effects on cosmic string
loops (see e.g. \cite{RHBloop}), and are
hence not robust.

Direct limits 
on the cosmic string tension $\mu$
can be obtained by looking for
specific signatures of individual long strings.
These are limits which are insensitive
to the parameters in the cosmic string scaling
solution.
Limits obtained from searching for the Kaiser-Stebbins
signature of a long string in CMB temperature
maps were derived in \cite{Lo,Smoot} using
WMAP data. The limits obtained were
weaker than the ones from (\ref{limit}).
However, the angular scale of the WMAP experiment
is too large to be able to effectively search for sharp features
in position space maps such as those predicted by
cosmic strings.
It was pointed out \cite{Fraisse} that the
string signatures for values of $G \mu$ somewhat
smaller than the limiting value of (\ref{limit}) should
be clearly visible in smaller angular scale
CMB anisotropy maps such as those provided
by the ACT \cite{ACT} and SPT \cite{SPT} experiments.
In recent work \cite{Amsel,Stewart,Rebecca1} it was
shown that limits up to an order of magnitude tighter
than (\ref{limit}) might be achievable using SPT
data. In the following, we will discuss direct signals
of wakes created by long strings in 21cm surveys. 

As first pointed out in \cite{Silk} and then
further discussed in \cite{Vach,Rees,SVBST,Charlton},
long strings moving perpendicular to the tangent
vector along the string give rise to ``wakes"
behind the string, i.e. in the plane spanned by
the tangent vector to the string and the velocity
vector. The wake arises as a consequence of
the geometry of space behind a long straight
string \cite{Vil2,Ruth} - space perpendicular
to the string is conical with a deficit angle given by
\be \label{deficit}
\alpha \, = \, 8 \pi G \mu \, .
\ee
From the point of view of an observer behind the string
(relative to the string velocity vector), it appears that matter
streaming by the string (from the point of view of the observer
traveling with the string) obtains a velocity kick of magnitude
\be
\delta v \, = \, 4 \pi G \mu v_s \gamma_s \, 
\sim \, 4 {\rm km/s} (G \mu)_6 v_s \gamma_s
\ee
towards the plane behind the string. In the above, $v_s$ is
the velocity of the string (in units of the speed of light),  
$\gamma_s$ is the corresponding relativistic gamma factor,
and $(G \mu)_6$ is the value of $G \mu$ in units of $10^{-6}$. This
leads to a wedge-shaped region behind the string
with twice the background density (see Figure 1).

\begin{figure}[htbp]
\includegraphics[height=5cm]{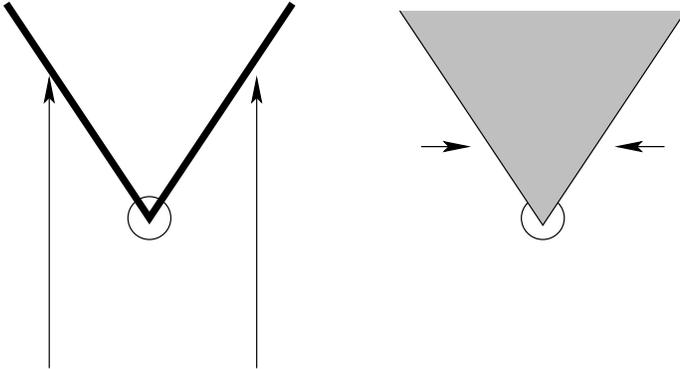}
\caption{Sketch of the mechanism by which a wake behind a moving
string is generated.  Consider a string perpendicular to the plane of
the graph moving straight downward. From the point of view of the
frame in which the string is at rest, matter is moving upwards, as
indicated with the arrows in the left panel. From the point of view
of an observer sitting behind the string (relative to the string motion)
matter flowing past the string receives a velocity kick towards the
plane determined by the direction of the string and the velocity
vector (right panel). This velocity kick towards the plane leads
to a wedge-shaped region behind the string with twice the
background density (the shaded region in the right panel).} \label{fig:1}
\end{figure}

Since the strings are relativistic, they generally move with a
velocity of the order of the speed of light. There will be
frequent intersections of strings. The long strings will  chop off
loops, and this leads to the conclusion that the string distribution
will be statistically independent on time scales larger than the
Hubble radius. Hence, to model the effects of strings we will
make use of a toy model introduced in \cite{Periv} and used
in most analytical work on cosmic strings and structure formation
since then: we divide the time interval between the time
of equal matter and radiation
and the current time $t_0$ into Hubble expansion time steps.
In each time step, we lay down a distribution of straight
string segments moving in randomly chosen directions with
velocities chosen at random between $0$ and $1$ (in units
of the speed of light). The centers and directions of these
string segments are also chosen randomly, and the string
density corresponds to $N$ strings per Hubble volume,
where $N$ is an integer which is of the order $1$ according
to the scaling of the string network. The distribution of
string segments is uncorrelated at different Hubble times.
Each string segment will generate a wake, and it is the
signal of one of these wakes  which
we will study in the following.
 
A string segment laid down at time $t_i$ will generate a wake 
whose dimensions at that time are the following:
\be \label{size}
c_1 t_i \, \times \, t_i v_s \gamma_s\, \times \, 4 \pi G \mu t_i v_s \gamma_s\, ,
\ee
where $c_1$ is a constant of order one. In the above, the first dimension is the length 
in direction of the string, the second is the depth. The fact that the string
wake has a finite depth is due to the causality constraints on
density fluctuations produced during a phase transition \cite{Traschen}:
The information about the formation of a string (and hence the information
about the existence of a deficit angle) cannot have propagated farther
than the horizon from the point of nucleation of the defect. It was
shown in \cite{Joao} that the deficit angle quite rapidly tends to zero
as the horizon is approached.
 
The overdense region in the wake will lead to the gravitational
accretion of matter above and below the wake towards the center
of the wake. In this way, the wake will grow in thickness. The
accretion of matter onto a cosmic string wake
was studied in \cite{SVBST,Leandros1} in the case of the dark matter 
being cold, and in \cite{Leandros1,Leandros2} in the case of the dark 
matter being hot.  We are interested in the case of cold dark matter.
As an aside, we mention that 
if the strings have lots of small-scale structure then
they will have an effective tension which is less than the effective
energy density \cite{wiggly}. This will lead to a local gravitational
attraction of matter towards the string, a smaller transverse
velocity, and hence to string filaments instead of wakes. The
gravitational accretion onto string filaments was studied in
\cite{Aguirre}. 

We will now review the computation of the width of the wake. 
We are interested in the distribution of baryons in the vicinity
of the string. However, for $t > t_{rec}$ the baryons and
cold dark matter feel the same gravitational attraction and
will thus behave in the same way - modulo thermal velocity
effects which will be discussed later - until the baryons undergo
a shock. Thus, for the moment we focus on the onset of
clustering of the cold dark matter. We are interested in 
wakes created after the time $t_{eq}$ of equal matter and radiation (there
is no gravitational clustering of cold dark matter before that
time). For times between $t_{eq}$ and recombination ($t_{rec}$),
the baryons are coupled to the radiation. However, for $t > t_{rec}$
the baryons will rapidly fall into the potential wells created by the cold
dark matter and thus, once again, it is legitimate to focus attention
on the clustering of the cold dark matter. Note that wakes produced
at the earliest time are the most numerous.
 
We consider a mass shell located at a comoving distance $q$ above the
wake. Its physical distance above the wake at time $t$ is
\be \label{height1}
w(q, t) \, = \, a(t) \bigl( q - \psi \bigr) \, ,
\ee
where $\psi$ is the co-moving perturbation induced by gravitational
accretion. If the wake is laid down at the initial time $t_i$ 
(with corresponding redshift $z_i $), then the
initial conditions for the cold dark matter fluctuation are 
\be \label{IC}
\psi(t_i) \, = \, {\dot{\psi}}(t_i) \, = \, 0 \, .
\ee
As a consequence of the initial wake planar overdensity $\sigma(t_i)$,
the co-moving displacement $\psi$ will begin to increase.
The clustering dynamics can be studied making use of 
the Zel'dovich approximation \cite{Zelap} in which the gravitational force is
treated in the Newtonian limit. As reviewed recently in
\cite{Danos3}, we obtain
\be \label{height2}
\psi(t) \, = \, \frac{18 \pi G}{5} \sigma(t_i) \bigl( \frac{t_i}{t_0} \bigr)^{2/3}
\bigl( \frac{t}{t_0} \bigr)^{2/3} t_0^2 \, ,
\ee
where $t_0$ is the present time and $\sigma(t_i)$ is the initial wake
planar density excess.

The mass shell with initial comoving distance $q$ above the wake ``turns
around" when ${\dot w}(q, t) = 0$. At time $t$, this occurs for a value
$q = q_{nl}(t)$ given by
\be \label{qnl}
q_{nl}(t) \, = \, \frac{24 \pi}{5} G \mu v_s \gamma_s (z_i  + 1)^{-1/2} t_0 \bigl( \frac{t}{t_i} \bigr)^{2/3}
\, .
\ee
It is easy to check that at the point of turnaround
\be
\psi(q_{nl}, t) \, = \, \frac{1}{2} q \, ,
\ee
and that hence the density inside the turnaround surface is twice the background
density.

Once a matter shell reaches its maximal distance $w_{max}$ 
from the center of the wake, it will start to collapse onto the wake.
The infall of matter will halt as the shell hits other streams
of matter. This will lead to a shock. In analogy with
what can be shown analytically in the context of the
spherical collapse model
(see e.g. \cite{Peebles}) we assume that the shock occurs
at approximately half the maximal distance. The 
hydrodynamical simulations of \cite{Sorn1} confirm the
applicability of this assumption. Since the
distance at turnaround is half the width the shell would
have without gravitational accretion onto the wake,
the shock occurs at a distance $1/4$ of that which the
matter shell would have under unperturbed Hubble
expansion. Hence, the average density inside the
wake is four times the background density, a result
which will be used several times in the computations
of the following section.

The evolution of the mass shell between turnaround and shock can be followed
using (\ref{height1}) and (\ref{height2}). The shock will occur when 
$w(q, t) = (1/2) w_{max}(q)$. A straightforward computation shows
that at this point the velocity is given by
\be \label{hresult}
{\dot{w}}(q, t) \, = \, - \frac{4 \pi}{5} G \mu v_s \gamma_s \bigl( \frac{z_i + 1}{z + 1} \bigr)^{1/2} \, .
\ee
It is this velocity which then determines the temperature inside the shocked 
region.

The shocks will lead to thermalization of the wake. The
gas temperature will be given by
\be \label{temp}
\frac{3}{2} k_B T \, = \, \frac{1}{2} m v^2 \, ,
\ee
where $m$ is the mass of a HI atom and
$v$ is the velocity of the in-falling particles
when they hit the shock, which is given by
\be \label{vel}
v \, = \, {\dot w}(q(t), t) \, ,
\ee
where $q(t)$ is the comoving distance of the shell which
starts to collapse at the time $t$.

Inserting the result (\ref{hresult}) into (\ref{vel}) and then
into (\ref{temp}) we obtain the following result for the
temperature $T_K$ of the HI atoms inside the wake
\bea \label{wakeT}
T_K \, &=& \, \frac{16 \pi^2}{75} (G \mu)^2 (v_s \gamma_s)^2 \frac{z_i + 1}{z + 1} k_B^{-1} m 
\nonumber \\
& \simeq &  [20~{\rm K}] (G \mu)_6^2 (v_s \gamma_s)^2 \frac{z_i + 1}{z + 1}  \, ,
\eea
where in the second line we have written the result in degrees K and
expressed $G \mu$ in units of $10^{-6}$ (and kept only one
significant figure in reporting the final number). The wake temperature
is largest for wakes produced at the earliest times since the initial wake density
is then the highest, and increases as time increases because more matter
has time to accrete.

Assuming values of $G \mu = 3 \times 10^{-7}$
(the current upper bound on the string tension), $z_i = 10^3$,
(close to the time of recombination),
$z + 1 = 30$ and $(v_s \gamma_s)^2 = 1/3$, Eq. (\ref{wakeT})
yields  $T_K \, \sim \, 20 K$.
This temperature is  smaller than the CMB temperature
$T_{\gamma} \simeq 82 K$ at this redshift. As we will see
in the following section, this leads to an absorption signal
in the 21cm radiation.

The formation of a cosmic string wake and the thermalization
which takes place inside the shocked region has been studied
in \cite{Sorn1} using an Eulerian hydro code \cite{Sorn2}
optimized to resolve shocks. The numerical simulations of
\cite{Sorn1} show that the density and temperature inside
the shocked region are indeed roughly uniform and that
the temperature obtained agrees with the values obtained
here using analytical approximations.

Looking for signals from cosmic strings in 21cm surveys is 
potentially more powerful than looking in large-scale optical redshift
surveys. This is because firstly the non-Gaussian signatures from
strings are more pronounced at higher redshifts, secondly
because we are directly looking at the distribution of the baryons,
and not just the distribution of stars, the latter being affected by
non-linear and gas dynamics, and thirdly because the distribution
of matter is more linear at higher redshifts and hence easier to
follow analytically. Compared to CMB and CMB polarization maps,
21cm surveys have the advantage of yielding three- rather than just
two-dimensional maps, maps thus containing much more
information. To our knowledge, there has been
little previous work on 21cm emission from strings.
Two exceptions are \cite{Wandelt} in which the angular
power spectrum of 21cm emission from a network of cosmic
strings was considered, and \cite{Aaron}, a recent study in which the
correlation between 21cm emission and CMB signals from
cosmic strings was studied. In contrast to these works, we are
looking for direct string signals in position space 21cm maps.

In the following section we will briefly summarize some key
features of 21cm cosmology and apply the equations
for the 21cm brightness temperature to the case of emission
from a string wake. 

\section{Cosmic Strings and 21cm Maps}

Neutral hydrogen is the dominant form of baryonic matter
in the ``dark ages", i.e. before star formation. Neutral
hydogen has a 21cm hyperfine transition line which
is excited if the hydrogen gas is at finite temperature.
Hence, we expect redshifted 21cm radiation to reach us
from all directions in the sky, and the intensity of this
radiation can tell us about the distribution of neutral
hydrogen in the universe, as a function of both angular
coordinates in the sky and redshift. Thus, in contrast
to the CMB, 21cm surveys can probe the three-dimensional
distribution of matter in the universe (see \cite{early}
for pioneering papers on the cosmology of the 21cm line
and \cite{Furlanetto} for an in-depth review).

Let us now consider the equation of radiative transfer along 
the line of sight through a hydrogen gas cloud of uniform 
temperature- in the case of interest to us this gas cloud is
the cosmic string wake. 
The brightness temperature $T_b(\nu)$ at an observed 
frequency $\nu$ due to 21cm emission is given by
\be \label{three1}
T_b(\nu) \, = \, T_S \bigl( 1 - e^{- \tau_{\nu}} \bigr) + T_{\gamma}(\nu) e^{- \tau_{\nu}} \, ,
\ee
where $T_S$ is the spin temperature,
$T_{\gamma}$ is the microwave radiation temperature,
and $\tau_{\nu}$ is the optical depth obtained by
integrating the absorption coefficient along the light ray
through the gas cloud. The frequency $\nu$ is the blue-shifted frequency
at the position of the cloud corresponding to the observed frequency
$\nu_o$.  The term proportional to $T_S$ is due to spontaneous emission, while
the term proportional to $T_{\gamma}$ is due to absorption and stimulated emission.

As explained in \cite{Furlanetto}, what is of observational interest is the 
comparison of the temperature coming from the hydrogen cloud with the 
``clear view'' of the 21 cm radiation from the CMB. 
\be  \label{three15}
\delta T_b(\nu) \, = \, \frac{ T_b(\nu)-T_{\gamma}(\nu) }{1+z} 
\, \approx \, \frac{ T_S-T_{\gamma}(\nu) }{1+z} \tau_\nu \, .
\ee
Note that the ``clear view'' of the CMB is hypothetical since even without a 
string wake's gas cloud the intergalactic medium is partly a less dense hydrogen 
gas. In the second part of the equation above we have expanded the exponential 
factor to linear order in the optical depth.

The spin temperature $T_S$ is defined as the relative number density of atoms 
in the hyperfine energy states through
$
{n_1}/{n_0} \, = \, 3 \exp({-T_{\star} / T_S} )\, .
$
Here $n_1$ and $n_0$ are the number densities of atoms in the
two hyperfine states, and $T_{\star}=E_{10}/k_B=0.068 {\rm K}$ is the temperature
corresponding to the energy splitting $E_{10}$ between these states. 

The spin temperature is determined solely by the temperature $T_K$ of the
gas in the wake, as long as UV scattering is negligible (which is true in our case). 
The relationship between spin and kinetic gas temperatures is expressed through 
the collision coefficients $x_c$ which describe
the rate of scattering among hydrogen atoms and electrons:
\be \label{three2}
T_S \, = \, \frac{ 1 + x_c}{1 + x_c T_{\gamma} / T_K} T_{\gamma} \, .
\ee
We will shortly discuss the numerical values of the $x_c$ for a particular case of interest. 

Combining (\ref{three15}) and (\ref{three2}) we find for the difference $\delta T_b(\nu)$
in brightness temperature induced by the interaction of the radiation with
neutral hydrogen is 
\be \label{three3}
\delta T_b(\nu) \, \simeq \, T_S \frac{x_c}{1 + x_c} \bigl( 1 - \frac{T_{\gamma}}{T_K} \bigr) 
\tau_{\nu} (1 + z)^{-1} \, ,
\ee
where the last factor represents the red-shifting of the temperature between the
time of emission and the present time. It is important to keep in mind that
the brightness temperature from a region of space without density perturbations
is nonzero. It is negative since the temperature of gas after redshift 200 is lower
than that of the CMB photons, the former red-shifting as $(z(t) + 1)^2$, the latter
as $z(t) + 1$.
The optical depth of a cloud of hydrogen is
\be \label{three5}
\tau_{\nu} \, = \, 
\frac{3 c^2 A_{10}}{4 \nu^2} \bigl( \frac{\hbar \nu_{10}}{k_B T_S} \bigr) \frac{N_{HI}}{4} \phi(\nu) \, ,
\ee
where $N_{HI}$ is the column density of HI. 

Up to this point, the analysis has been general. Let us now specialize to the case
where the gas cloud is the gas inside the cosmic string wake. In this case, 
the column density is the hydrogen 
number density $n_{HI}^{wake}$ times the length that the light ray traversed 
in the cloud. This length will depend on the width $w$ of the wake and the angle 
$\theta$ that the light ray makes relative to the vertical to the wake so that 
\be
N_{HI} \, = \, \frac{2 n_{HI}^{wake} w}{\cos\theta} \, .
\ee
The factor of $2$ results since $w$ is the width of the wake from the center.

The line profile $\phi(\nu)$ is due to broadening of the emission line.
This broadening reflects the fact that not all photons resulting from the hyperfine
transition will leave the gas cloud at the same frequency. Frequency differences
are in general due to thermal motion, bulk motion and pressure effects.
Since the pressures we are considering are small by astrophysical standards,
pressure broadening is negligible. Since the gas temperature inside the wake
is not much larger than the CMB temperature, thermal broadening will not be
important, either. This leaves us with the effects of bulk motion, more specifically
the expansion of the wake in planar directions. The line profile is normalized
such that the integral of it over frequency is unity.

The origin of the line broadening due to the expansion of the wake is illustrated
in Figure 2. Let us consider a point on the wake for which the photons travel to us
at an angle which is not orthogonal to the plane of the wake. Then, relative
to photons emitted at the central point on the photon path, photons emitted
from the highest point and the lowest point obtain a relative Doppler shift of
\be \label{three7}
\frac{\delta \nu}{\nu} \, = \, 2 {\rm{sin}}(\theta) \tan{\theta} \frac{H w}{c} \, ,
\ee
where $H$ is the expansion rate of space and $w$ is the wake width
computed in the previous section, both evaluated at the redshift $z$ when the
photons are emitted.  The angle $\theta$ is indicated in Figure 2. It is
the angle relative to the vertical to the wake.
As a consequence of the normalization of $\phi(\nu)$
we hence find
\be \label{three8}
\phi(\nu) \, = \, \frac{1}{\delta \nu} \,\,\, \rm{for} \,\,\, 
\nu \, \epsilon \, [\nu_{10} - \frac{\delta \nu}{2}, \nu_{10} + \frac{\delta \nu}{2}] \, ,
\ee
and $\phi(\nu) = 0$ otherwise. 

In addition to its role in determining the brightness
temperature, the frequency shift $\delta \nu$ is important since
it determines the width of the 21cm signal of strings in the
redshift direction and is hence central to the issue of observability
of the signal. From (\ref{three7}) and (\ref{qnl}) (and
setting the $2 sin{\theta} tan{\theta}$ factor to one) we find that
\bea \label{freqwidth}
\frac{\delta \nu}{\nu} \, &=& \, \frac{24 \pi}{15} G \mu v_s \gamma_s 
\bigl( z_i  + 1 \bigl)^{1/2} \bigl( z(t) + 1 \bigr)^{-1/2} \nonumber \\
& \simeq & \, 3 \times 10^{-5} (G \mu)_6 (v_s \gamma_s) \, ,
\eea
where in the second line we have used the cosmological 
parameters mentioned at the end of the introductory section and 
inserted the representative redshifts
$z_i  + 1 = 10^3$ and $z(t) + 1 = 30$.

\begin{figure}
\includegraphics[height=5cm]{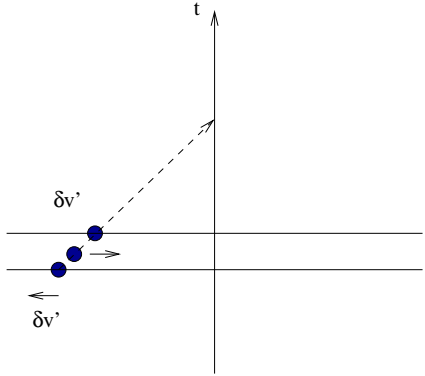}
\caption{Photons reaching us from a particular direction given by the angle $\theta$ 
to the vertical to the wake are emitted at different points in the wake. Three such points
are indicated in the Figure - the central point and two points on the bottom and top of
the wake, respectively. Since the wake is undergoing Hubble expansion in its
planar directions, there is a relative Doppler frequency shift $\delta \nu$ between
the 21cm photons from the top and the center of the wake.} \label{fig:2}
\end{figure}

Taking the formula (\ref{three3}) for the brightness temperature, inserting
the results (\ref{three5}) for the optical depth and
(\ref{three7}) and (\ref{three8}) for the line profile, we get
\bea \label{result1}
\delta T_b(\nu) \, &=& \, 2 \frac{x_c}{1 + x_c} \bigl( 1 - \frac{T_{\gamma}}{T_K} \bigr) 
\frac{3 c^3 A_{10} \hbar }{16 \nu_{10}^2 k_B H_0 \Omega_m^{1/2}}
\nonumber \\
& & \,\, \times n_{HI}^{bg}(t_0)\frac{n_{HI}^{wake}(t_0)}{n_{HI}^{bg}(t_0)} 
\nonumber \\
& & \,\, \times (2 {\rm{sin}}^2(\theta))^{-1} (1 + z)^{1/2} \, ,
\eea
where $\Omega_m$ is the fraction of the critical energy density which
is in the form of matter.
Here the ratio of the density inside the wake to the background density, 
$n_{HI}^{wake}/n_{HI}^{bg}$, is approximately 4. 
We have rescaled the Hubble constant and the background HI density
to its current values $H_0$ and $n_{HI}^{bg}(t_0)$, respectively, and made
use of $H(z) = H_0 \Omega_m^{1/2} (1 + z)^{3/2}$ in the re-scaling.
Note that the width of the wake has cancelled out between the HI column density
and the line profile. The width of the wake, however, has not disappeared
completely from the calculation since it determines the wake temperature
$T_K$, and since it yields the width of the signal in redshift direction.

Taking the values $A_{10} = 2.85 \times 10^{-15}~{\rm s}^{-1}$, $T_{\star} = 0.068~{\rm K}$, $H_0=73~{\rm km~s}^{-1}~{\rm Mpc}^{-1}$, 
$\nu_{10}=1420$~MHz,   $\Omega_b=0.042$, $\Omega_m=0.26$,
$2\sin^2\theta=1$, eq.~(\ref{result1}) becomes
\be \label{result2}
\delta T_b(\nu) \, = \, [0.07~{\rm K}] \frac{x_c}{1 + x_c} \bigl( 1 - \frac{T_{\gamma}}{T_K} \bigr) 
(1 + z)^{1/2}  \, .
\ee

The collision coefficient $x_c$ is dominated by the hydrogen-hydrogen collisions 
because of the very low fraction of free electrons. It is given by \cite{Zygelman}
\be
x_c \, = \, \frac{n \kappa_{10}^{HH}}{A_{10}} \frac{T_{\star}}{T_{\gamma}} \, .
\ee
where $\kappa_{10}^{HH}$ is the de-excitation cross section and is approximately 
$2.7 \times 10^{-11} \rm{cm^3 s^{-1}}$ at a wake gas temperature of 
$T_K = 20 K$ \cite{Zygelman}. 
This temperature corresponds to the parameter values we took at the end of Section~2, i.e. 
$(G \mu)_6 = 0.3$, $(v_s \gamma_s)^2 = 1/3$, $z_i = 10^3$ and
$z + 1 = 30$.
Remembering that the density $n$ inside the wake is four times the
background density $n_{bg} = 1.9 \times 10^{-7} {\rm{cm}}^{-3} (1 + z)^3$ we 
find for a redshift $1 + z = 30$ that
$x_c \, \simeq \, 0.16$ and hence (from (\ref{result2}))
$\delta T_b(\nu) \, \simeq \,  - 160~{\rm mK}$. For a formation
redshift $z_i = z_{eq} = 3200$ corresponding to the time of equal matter and radiation
the corresponding temperature is $\delta T_b(\mu) \, \simeq \, -15~{\rm mK}$.
Note that the dependence of the above result on the cosmic string tension
$\mu$ enters only through the wake temperature. Note also that in the large
$G \mu$ limit ($G \mu \gg 10^{-6}$), the brightness temperature 
approaches a constant value of close to $400 {\rm mK}$. 

There is a critical value of the string tension (which depends on the
redshift) at which the cosmic string signal changes from emission
to absorption. This value is determined by $T_K = T_{\gamma}$
which yields
\be \label{crit1}
(G \mu)_6^2 \, \simeq \,  0.1 (v_s \gamma_s)^{-2} \, \frac{(z + 1)^2}{z_i + 1} \,
\simeq \, 0.4 \, \frac{(z + 1)^2}{z_i + 1} \, ,
\ee
for the value $(v_s \gamma_s)^2 = 1/3$ which we are using throughout,
and this corresponds to $(G \mu)_6 \, \simeq \, 0.6$ if we
use $(z_i + 1) = 10^3$ and insert our reference redshift of $z + 1 = 30$.
At a redshift of $z + 1 = 20$ the critical value is $(G \mu)_6 \, \simeq \, 
0.4$. For a formation time corresponding to the redshift of equal matter
and radiation the critical value of $G \mu$ is a factor of two lower.

For lower values of the tension the cosmic string signal in the 21cm maps shifts from
an emission signal to an absorption signal. However, two effects must be taken
into account when discussing the
visibility of the string signal. The first is the fact that
at very small values of $G \mu$ our calculation of the wake temperature breaks down. 
This occurs when the resulting wake temperature computed
in (\ref{wakeT}) is smaller than the gas temperature obtained by taking
the background gas temperature and computing its increase under
adiabatic compression to the overdensity of the wake. In this case it is no longer
justified to take the initial conditions (\ref{IC}). At redshifts $z$ below 
150 the cosmic gas is cooling adiabatically as 
\be \label{gasT}
T_g \, = \, 0.02 {\rm K} (1+z)^2 
\ee
because at this point Compton heating of the gas by the CMB is negligible~\cite{SSS}.
Setting $T_K = 2.5 \times T_g$ (the factor $2.5$ being due to the fact that
for adiabatic compression of a mono-atomic gas by a factor of $4$ in volume, one
expects a temperature increase by a factor of $4^{2/3} \sim 2.5$) 
yields (for our standard values of
$1 + z_i = 10^3$ and $(v_s \gamma_s)^2 = 1/3$)
\be \label{crit2}
(G \mu)_6 \, \simeq \, 3 \times 10^{-3} (1 + z)^{3/2}
\ee
which equals $(G \mu)_6 = 0.5$ at a redshift of $1 + z = 30$. For
a formation redshift $z_i = z_{eq}$ the factor $3$ in (\ref{crit2}) is
replaced by $2$.

For values of  $G \mu$ smaller than the one given in (\ref{crit2}), the
effects of thermal pressure start to become important, and this
will effect both the density and temperature structure in the
wake.

Our results are illustrated in Figure 3 which shows a comparison
of the temperatures relevant to the above discussion and their
dependence on redshift. The vertical axis is the temperature
axis, the horizontal axis is inverse redshift. The magenta (dashed)
line shows the CMB temperature $T_{\gamma}$, the orange
(dotted) line is the gas temperature in a wake $2.5 T_g$ after
adiabatic contraction. The two solid lines with the positive slope
represent $T_K$ for two different values of $G \mu$. The upper
curve is for $(G \mu)_6 = 1$, the lower curve for $(G \mu)_6 = 0.3$.
Along a fixed $G \mu$ curve, the 21cm signal of a cosmic string
is an emission signal above the $T_{\gamma}$ curve and an
absorption signal below it. Above the $2.5 T_g$ curve, the 
temperature of the gas inside the wake is well approximated by 
the equation (\ref{wakeT}), below it the initial thermal gas temperature
effects dominate and the wake temperature will follow the
$2.5 T_g$ curve. The wake temperature curves are for a
formation redshift of $z_i  + 1 = 10^3$.

\begin{figure}
\includegraphics[height=5cm]{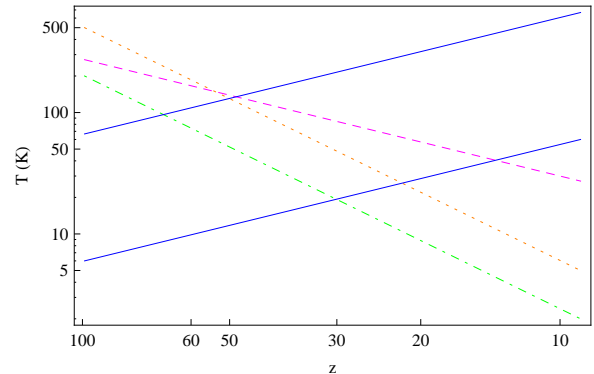}
\caption{The redshift dependence of the relevant temperatures. The majenta
dashed line is the CMB temperature $T_{\gamma}$, the orange
dotted line is $2.5 T_g$, the gas temperature after adiabatic contraction
by a factor 4 in volume, and the green (dot-dashed) line is
the $T_g$ curve. The two solid curves with positive slope represent
$T_K$ for two different values of $G \mu$, $(G \mu)_6 = 1$ in the case
of the upper curve and $(G \mu)_6 = 0.3$ in the case of the lower
curve (in both cases for a formation redshift $z_i  + 1 = 10^3$). 
Following one of the solid curves, we see that the 21cm string
signal is an emission signal above the majenta curve and an
absorption signal below it. The wake temperature is given by $T_K$
only above the orange curve. For higher redshifts, the initial gas temperature
dominates the final temperature of the gas inside the wake, and the
latter follows the orange curve. Once $T_K$ drops below the
green curve, there will no longer be any shock and our effect
disappears.} \label{fig:3}
\end{figure}

The second issue is that the
string signal must be compared to what would be seen if the wake region were
replaced by unclustered neutral gas. In this case, the absorption signal
has a temperature given by (\ref{result2}) with $T_K$ replaced by $T_g$
and the collision coefficient computed with $T_g$ instead of $T_K$.
Due to the overdensity of the wake, the signal of a string wake should
persist. The overdensity in the string wake by a factor of 4 will
lead to an enhancement in the magnitude of the brightness temperature
by a factor of 16 compared to what would be seen if the wake region were
replaced by unclustered gas. One factor of 4 comes from the fact that
the collision coefficient is proportional to the gas density, the second
from the fact that the optical depth is also proportional to the
density. In addition, the temperature ratios in
(\ref{result2}) are different. 

However, for values of $G \mu$ so small that $T_K \ll T_g$ there
will no longer be any shock and hence no well-defined wake region
with overdensity 4. This will occur at a value of $G \mu$ which 
is smaller than the limit in (\ref{crit2}) by a factor of $\sqrt{2.5}$.
Note also that the thickness in redshift space (discussed
below) of the signal decreases as $G \mu$ decreases, and hence
improved sensitivity will be required to detect the signal.

The bottom line of the above analysis is that at a 
redshift of $z + 1 = 30$, strings with tensions
below the current observational limit are predicted to be visible 
in absorption in 21cm surveys. Strings with larger tensions
would yield an emission signal. The critical value (\ref{crit1})
of $G \mu$ at which
the emission signal turns into an absorption signal decreases as $z$ decreases,
so that strings with tensions at the current observational limit would become
visible in emission at a redshift of below 20.
However, the value of $z$ cannot be smaller than that corresponding to reionization 
since we have not considered how UV scattering affects the spin temperature.

The 21cm signature of a cosmic string wake has a distinctive shape in redshift
space. The string signal will be a wedge-like region of extra
21cm emission. The signal will be wide in the two angular directions,
and narrow in redshift direction \footnote{The signal at a fixed
redshift would be extended only in one angular direction.}.
The projection of this region onto the plane corresponding
to the two angles in the sky (we are working in the limit of
small angles and thus can use the flat sky approximation) corresponds
to the projection of the wake onto the observer's past light cone. The
orientation of the wedge in redshift space is given by the orientation
of the wake relative to the observer's light cone. This is illustrated
in Figure 4. The left panel is a space-time sketch of the geometry (with
two spatial directions suppressed). The wake is created by a
string segment which starts out at time
$t_i$ at the position $x_1$ and which at time $2 t_i$ (roughly one
Hubble time step later) has moved to the position $x_2$ (the arrow on the
segment connecting the two events gives the direction of the
velocity of the string segment. For the orientation of the string
chosen, the past light cone intersects the ``back'' of the string wake
at a later time than the front. Hence, the 21cm signal has a larger
red-shift for photons from the tip of the wake than from the back side
of the wake. The width of the wake vanishes at the tip and increases
towards the back. Hence, in redshift space the region of extra 21cm
emission has the orientation sketched on the right panel of Figure 4.
The planar dimensions of the region of extra emission
are in the angular directions, and their sizes are the angles corresponding to
the comoving area (see (\ref{size}))
\be
c_1 t_i (z_i + 1) \times t_i v_s \gamma_s (z_i + 1) \, .
\ee
The amplitude of the emission signal depends on the orientation of the
wake relative to us (through the redshift when our past light cone
intersects the various parts of the wake). This is not correlated with
the direction of string motion. The width in redshift space, on the
other hand, depends on the emission point on the wake - the width
is larger in the back and approaches zero at the tip, as given by
$\delta \nu$ (see \ref{three7}). 


\begin{figure}
\includegraphics[height=4.7cm]{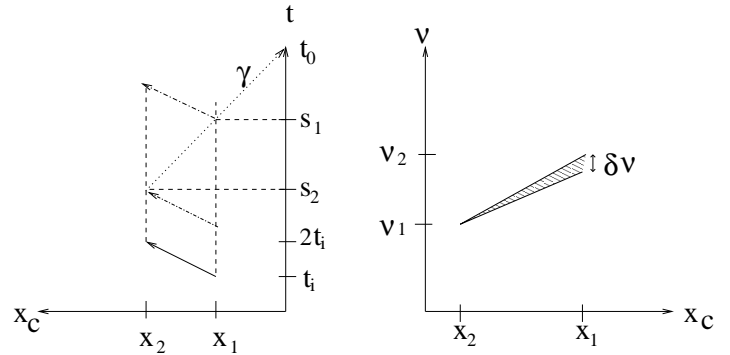}
\caption{Geometry of the 21cm signal of a cosmic string wake. The
left panel is a sketch of the geometry of the wake in space-time -
vertical axis denoting time, the horizontal axis one direction of
space. The string segment producing the wake is born at time $t_i$ 
and travels in the direction of the arrow, ending at the position 
$x_2$ at the time $2 t_i$. The past light cone of the observer
at the time $t_0$ intersects the tip of the wake at the time $s_2$,
the back of the wake at time $s_1$. These times are in general
different. Hence, the 21cm radiation from different parts of the
wake is observed at different red-shifts. The resulting angle-redshift
signal of the string wake shown in the left panel is illustrated
in the right panel, where the horizontal axis is the same spatial
coordinate as in the left panel, but the vertical axis is the
redshift of the 21cm radiation signal. The wedge in 21cm has vanishing
thickness at the tip of the wedge, and thickness given by $\delta \nu$
at the back side.} \label{fig:4}
\end{figure}


\section{Conclusions and Discussion}

In this paper we have calculated the 21cm signal of a single cosmic
string wake. We found that wakes leave a characteristic signal
in 21cm surveys - wedge shaped regions of extra emission or
absorption, depending on the value of the string tension.
There is an emission signal as long as the
string tension exceeds the critical value given by (\ref{crit1})
and as the tension increases the temperature approaches
the value given by $200  {\rm mK}$, a value larger
by almost two orders of magnitude compared to the backgrounds
from standard cosmology sources which the Square Kilometer
Array (SKA) telescope project is designed to be sensitive
to \cite{Carilli}. For values of $G \mu$
below the critical value given by (\ref{crit1}), the 
cosmic string wake signal changes to an absorption
signal. This will persist down to a value of $G \mu$
a factor of $\sqrt{2.5}$ smaller than the value
given in (\ref{crit2}), at which point the
material which is gravitationally attracted to the
wake will no longer undergo a shock and hence our
analysis ceases to be applicable. The width of the wedge
in redshift space depends on the
string tension via (\ref{three7}). The width increases
from zero at the tip of the string to the value given
by (\ref{three7}) at the midpoint of the wake.

The sensitivity of 21cm surveys to cosmic strings is
best at the lowest redshifts sufficiently higher than the
redshift of reionization. This is due to the fact that
the string wakes keep increasing in width as a function
of time, leading to an increasing temperature of the gas
in the wake and hence to a higher 21cm signal. At the
same time, the wakes become more stable towards
thermal disruption.

As mentioned above, the brightness 
temperature shift which we predict will
be large compared to that from the surrounding
intergalactic medium. The fractional frequency
width of the signal is given by (\ref{freqwidth}) and
is near the limit of the frequency resolution of radio
telescopes for values of $G \mu$ close to the
current upper bound. For example, the Square Kilometer
Array will have a fractional spectral resolution of the 
order \cite{Carilli} of $10^{-4}$. In both angular directions,
the signal will cover a rectangle whose side length is
given by the comoving
Hubble radius at the redshift $z_i $, which for redshifts
close to recombination corresponds to a scale of about
one degree. Thus, the signal of cosmic strings which we
predict is clearly a possible target for future radio telescopes.
For example, the angular resolution of the SKA is designed
to be smaller than $0.1$ arcsec, and the frequency range of the
SKA will allow the detection of the 21cm signal up to a
redshift of 20.

We have focused on the signal of a single wake laid down
at a time $t > t_{eq}$. The reader may worry that the signal
of such a wake is masked by the ``noise'' due to wakes laid
down at earlier times. Even though the baryons will stream
out of these wakes due to their coupling with the photons, the
dark matter wakes persist. For $t < t_{eq}$ the dark matter
wakes do not grow in thickness until $t = t_{eq}$. After that,
the thickness will begin to grow, and at $t > t_{rec}$ the 
baryons will start to fall into the dark matter potential
wells. However, the width of such wakes (averaged over the
length of the wake) is smaller than that of wakes at $t = t_{eq}$.
In addition, due to their large number, on length scales of wakes 
laid down at $t \geq t_{eq}$, the earlier wakes will act like
Gaussian noise. The coherent signature in position space
of such a ``late'' wake can be picked out of the Gaussian
noise even if the amplitude of the Gaussian noise is comparable to
or slightly larger than the amplitude of the signal (as measured
in terms of the contribution to the power spectrum). This
has been studied in the context of picking out the signatures
of late strings in CMB temperature maps in \cite{Amsel,Stewart,Rebecca1},
and we expect similar conclusions to hold here.

Let us end with a brief comparison of our work with that of
\cite{Wandelt} and \cite{Aaron} who also considered 21cm signals of cosmic
strings. What sets our work apart is that we focus on the specific 
position space signature of wakes rather than on the (Fourier space)
power spectrum. In computing a power spectrum, one loses the
information about the non-Gaussianities in the distribution of
strings. It is these non-Gaussianities which most clearly distinguish
the predictions of a string model from a model with only Gaussian
fluctuations. Therefore, the sensitivity to the presence of cosmic
strings will be much higher in a study like ours compared to what
can be achieved by only computing power spectra.

\begin{acknowledgments} 
 
This work is supported in part by a NSERC Discovery Grant, by funds from the 
CRC Program, by the FQRNT Programme de recherche 
pour les enseignants de coll\`ege, and by a Killam Research Fellowship awarded to R.B.
We wish to thank U.-L. Pen, J. Magueijo, R. Rutledge and in particular 
Y. Wang for useful discussions.

\end{acknowledgments}

\end{document}